\documentclass[sigconf, anonymous=false, nonacm]{acmart}

\AtBeginDocument{%
  }

\setcopyright{acmlicensed}
\copyrightyear{2018}
\acmYear{2018}
\acmDOI{XXXXXXX.XXXXXXX}
\acmConference[Conference acronym 'XX]{Make sure to enter the correct
  conference title from your rights confirmation email}{June 03--05,
  2018}{Woodstock, NY}
\acmISBN{978-1-4503-XXXX-X/2018/06}

\usepackage[utf8]{inputenc} 
\usepackage[T1]{fontenc}    
\usepackage{hyperref}       
\usepackage{url}            
\usepackage{booktabs}       
\usepackage{amsfonts}       
\usepackage{enumitem}       
\usepackage{nicefrac}       
\usepackage{microtype}      
\usepackage{xcolor}         
\usepackage{graphicx} 
\usepackage{multirow}
\usepackage{hyperref}
\usepackage{amsmath}
\usepackage{geometry}
\usepackage{xcolor} 
\usepackage{colortbl} 
\usepackage{booktabs} 
\usepackage{todonotes}

\usepackage{subfig}
\usepackage{balance}

\begin{document}

\title{ATLAS: Efficient Atom Rearrangement for Defect-Free Neutral-Atom Quantum Arrays Under Transport Loss}

\author{Otto Savola}
\email{otto.savola@aalto.fi}
\affiliation{%
  \institution{Aalto University}
  \city{Espoo}
  \country{Finland}
}

\author{Alexandru Paler}
\email{alexandru.paler@aalto.fi}
\affiliation{%
  \institution{Aalto University}
  \city{Espoo}
  \country{Finland}
}

\renewcommand{\shortauthors}{Savola et al.}

\begin{abstract}
Neutral-atom quantum computers encode qubits in individually trapped atoms arranged in optical lattices. Achieving defect-free atom configurations is essential for high-fidelity quantum gates and scalable error correction, yet stochastic loading and atom loss during rearrangement hinder reliable large-scale assembly. 
This work presents ATLAS, an open-source atom transport algorithm that efficiently converts a randomly loaded $W \times W$ lattice into a defect-free $L \times L$ subarray while accounting for realistic physical constraints, including finite acceleration, transfer time, and per-move loss probability. 
In the planning phase, optimal batches of parallel moves are computed on a lossless virtual array; during execution, these moves are replayed under probabilistic atom loss to maximize the expected number of retained atoms. 
Monte Carlo simulations across lattice sizes $W=10$--$100$, loading probabilities $p_{\mathrm{occ}}=0.5$--$0.9$, and loss rates $p_{\mathrm{loss}}=0$--$0.05$ demonstrate fill rates above $99\%$ within six iterations and over $90\%$ atom retention at low loss. 
The algorithm achieves sublinear move scaling ($\propto M^{0.55}$) and linear growth of required initial size with target dimension, outperforming prior methods in robustness and scalability---offering a practical path toward larger neutral-atom quantum arrays.
\end{abstract}



\keywords{neutral atoms, atom rearrangement, defect-free}


\maketitle

\section{Introduction}

Quantum computing has the potential to revolutionize many fields, such as medicine~\cite{durant_primer_2024} and computing itself~\cite{preskill_quantum_2018}, by leveraging superposition and entanglement. 

However, the practical usage is still limited since the quantum computing hardware is not yet sufficiently developed. One of the most promising approaches for hardware is utilizing neutral atoms as qubits~\cite{henriet2020quantum,bluvstein2025fault}. Compared to superconducting technology, neutral atom technology has several advantages. It has longer coherence times, allowing more operations to be done with the qubit~\cite{Qompose_Silver}. Using optical traps, neutral atoms can be organized in arrays of various geometries that may hold even thousands of atoms~\cite{pause_supercharged_2024}. Additionally, atoms of the same isotope are naturally identical, making the individual calibration of qubits unnecessary~\cite{wintersperger_neutral_2023}. Lastly, it offers a possibility of all-to-all connectivity between qubits~\cite{wintersperger_neutral_2023}. This can be achieved by physically moving atoms close to each other and exciting them to Rydberg states using laser pulses. With these characteristics, neutral atoms contain a foundation for scalable quantum computing. 

Despite these advantages, a key challenge still remains in neutral atom quantum computing: ensuring defect-free atom arrays. Attempting to perform quantum operations on missing qubits would lead to errors, making defect-free arrangement necessary before computation could begin, especially for current error correction codes~\cite{lidar2013quantum}. Addressing this challenge of rearranging atoms into a defect-free lattice enables the development of more reliable neutral atom quantum computers.

Atom rearrangement in neutral-atom systems closely mirrors design-automation tasks in quantum circuit compilation~\cite{schmid2024computational, zulehner2018efficient, tan2024compiling}. In both settings, one must transform an initial, imperfect resource layout into an optimized configuration that satisfies hardware constraints while minimizing costly operations. Rearranging atoms involves routing mobile qubits through a constrained 2D space, avoiding collisions, preserving ordering rules, and optimizing movement distance—all analogous to qubit routing, placement, and swap scheduling in quantum circuit design tools.

\section{Background}

\begin{figure}
    \centering
    \subfloat[]{\includegraphics[width=0.25\linewidth]{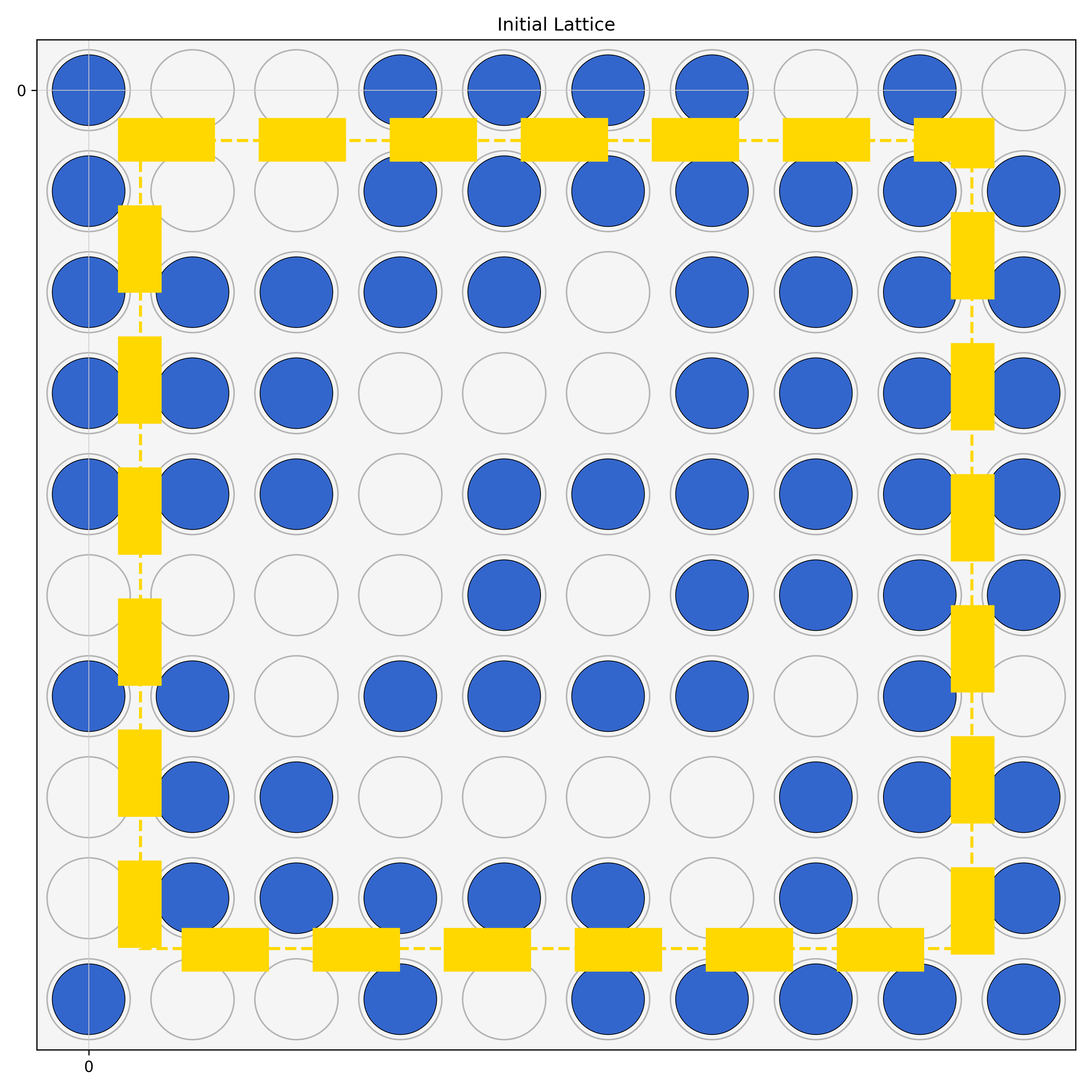}\label{fig:10x10_initial}}
    \hfil
    \subfloat[]{\includegraphics[width=0.25\linewidth]{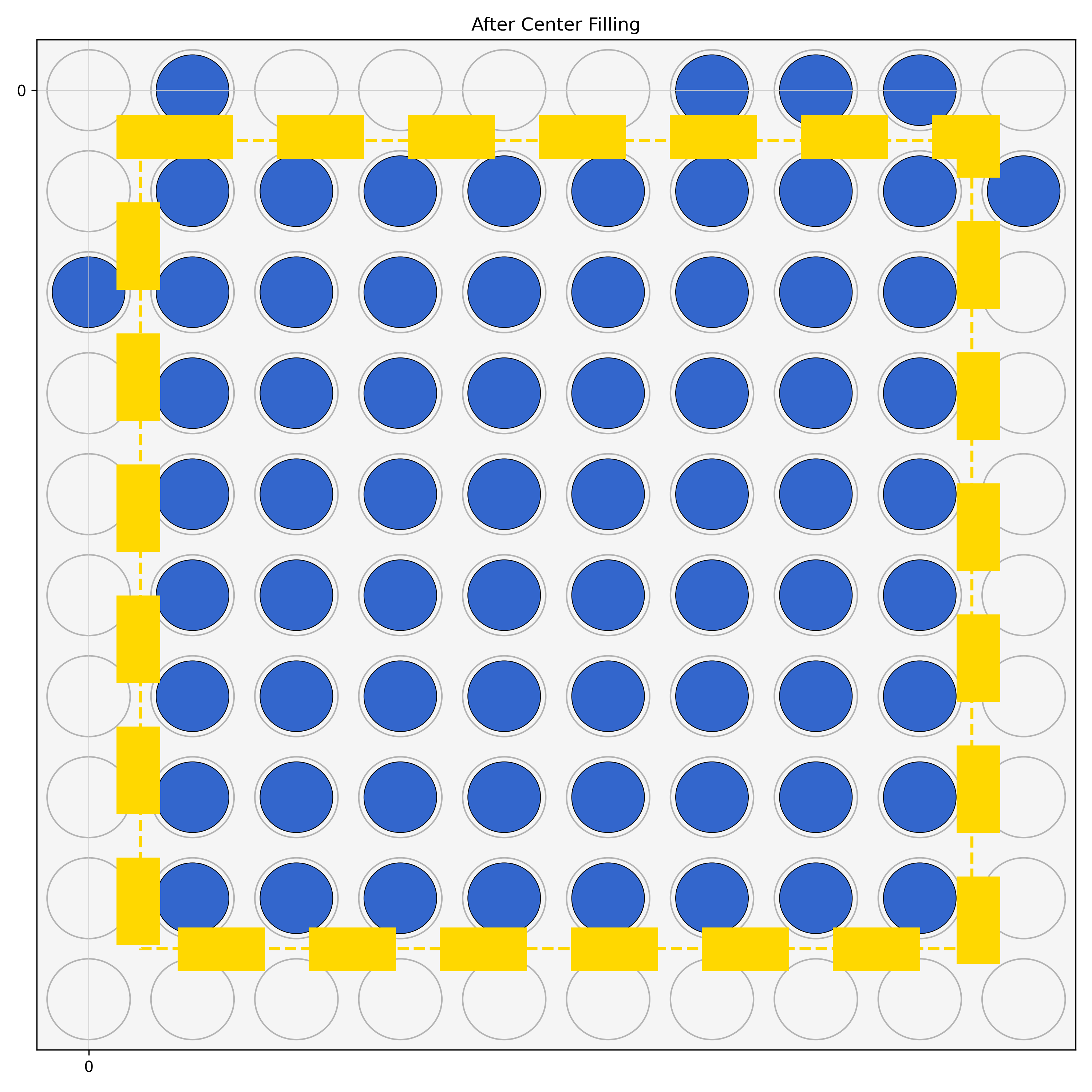}\label{fig:10x10_final}}
    \caption{Rearrangement of $10 \times 10$ neutral atom array. Blue disks indicate occupied traps (atoms) and grey rings mark all SLM trap sites. \textbf{(a)} The initial, probabilistically loaded configuration ($p_{occ} \approx 0.7$), with randomly missing atoms. \textbf{(b)} The defect-free square lattice achieved after running the assembly algorithm.}
    \label{fig:before_and_after}
\end{figure}

The basic idea of neutral atom quantum computing is to cool the atoms (e.g., rubidium-87 or cesium) down to near absolute zero temperature. Cooling is the same as slowing the atoms since the kinetic energy of atoms defines heat. After slowing the atoms, they are usually transferred into stationary traps that hold the atoms in fixed positions. However, since the atom loading process is inherently probabilistic, not all trap sites are occupied (see Fig.~\ref{fig:10x10_initial}). A uniform region of atoms is most likely not achieved, so the atoms are shuttled using mobile traps to form a defect-free uniform shape (see Fig.~\ref{fig:10x10_final}). Only then can a quantum algorithm be performed.

To rearrange the atoms to a defect-free configuration, mobile tweezers, typically generated using acousto-optic deflectors (AODs), are utilized~\cite{barredo_atom-by-atom_2016}. The rearrangement process usually begins by taking a fluorescence image of the initial atom distribution within the traps~\cite{wintersperger_neutral_2023}. In the resulting image, the atoms appear bright whereas empty traps appear dark. Once the initial ordering has been established, atoms are transferred between traps using mobile tweezers to create a defect-free arrangement. For the purpose of this work, the initial lattice is formed by spatial light modulator (SLM) traps, which are programmable optical device that shapes laser light to generate large, static 2D arrays of trapping sites for neutral atoms. The mobile traps are generated by two perpendicular 1D AODs. Thus, all the AOD-generated traps can only be moved in columns or rows, not individually. Moreover, to avoid heating and loss of the atoms when moving them, the columns AOD rows and columns cannot cross each other~\cite{bluvstein_quantum_2022}.

\textbf{Observation:} The maximum acceleration that can be applied to a trapped atom is one of the primary physical constraints. Experiments have shown that atoms start to escape when acceleration exceeds 2750 $m/s^2$ under constant jerk ~\cite{bluvstein_quantum_2022}. We simplified this by implementing a trapezoidal velocity profile. In our trapezoidal model, atoms first accelerate at a constant 2750 $m/s^2$ until reaching 0.13 $m/s$~\cite{tian_parallel_2023}, then continue at that speed until decelerating at the same rate to reach zero velocity.

\textbf{Observation:} Transferring an atom from a static trap into a mobile trap takes time: 60 $\mu$s~\cite{tian_parallel_2023} and the same amount of time for transferring the atom from the mobile trap back to a static trap. Therefore, the total time to move an atom from a trap at point a into a trap at point b will take: $2 \times 60\mu s + \mathrm{movement~time}$.

\textbf{Observation:} The static optical trap sites cannot be placed too close to each other. The minimum spacing is determined by the diffraction limit of the optics and the pixel size of the SLM~\cite{nogrette_single-atom_2014}. We use a distance of 5 $\mu m$ between our static SLM traps~\cite{tian_parallel_2023} (4.49 $\mu m$).

Even under these constraints, some atom losses will happen during the transportation due to imperfect transfers between traps and sudden changes in potential as the atoms are hovering over the static traps~\cite{tian_parallel_2023}.

\subsection{Contribution}

This work introduces ATLAS, a scalable and loss-aware algorithm for
assembling defect-free neutral-atom arrays (Section~\ref{sec:methods}). Its main contributions are: 1) A loss-aware target-sizing method that predicts achievable array dimensions under finite transport loss; 2) A physically compliant parallel-transport scheme that respects AOD motion constraints and maximizes simultaneous moves; 3) A two-phase plan--then--execute framework that enables efficient computation while modeling realistic loss during transport; 4) Extensive simulations (Section~\ref{sec:res}) showing high fill rates, strong retention, sublinear move scaling, and linear initial-size requirements; 5) A parallelization refinement (Section~\ref{sec:par}) that further reduces move scaling matching leading multitweezer methods.

\subsection{Problem Statement}

The initial lattice is a two-dimensional matrix $M$ of dimensions $W \times W$, where $M_{i,j} = 1$ indicates an atom being present at position  $(i,j)$, and $M_{i,j} = 0$ indicates no atom is present at position $(i,j)$. For each position, $M_{i,j}$ an atom is loaded with a probability $p_{occ}$. After initial loading, the expected number of atoms in the lattice is given by $I = W \times W \times p_{occ}$.

The pioneering work for rearranging the atoms is~\cite{Atom-by-atom_assembly} and~\cite{barredo_atom-by-atom_2016}, where the prior one is rearranging the atoms in a 1D array and the latter is in 2D. However, in both of these methods atom moves are executed sequentially, so at most one defect can be filled in one move. More recent works~\cite{cimring_efficient_2023,tian_parallel_2023, wang_accelerating_2023} focus specifically on parallel rearrangement using AODs to move multiple atoms simultaneously. These enhance scalability significantly by reducing the moves executed. Even better scalability has been demonstrated utilizing ultra-high speed SLMs, which introduce fully parallel atom transport~\cite{knottnerus_parallel_2025, lin_ai-enabled_2024}. While offering a promising, near-constant scaling compared to array size, it is not realistic with the current apparatus.

Key performance metrics for evaluating the rearrangement are fill rate (Section~\ref{sec:fill}), retention rate (Section~\ref{sec:ret}) and time (Section~\ref{sec:time}).

\textbf{Definition:} The final fill rate $\eta = \frac{{F}}{N} $ quantifies the density of atoms in the target zone (e.g. yellow in Fig.~\ref{fig:before_and_after}), where F is the number of atoms in the target zone at the termination and $N$ is the number of traps in the target zone.

\textbf{Definition:} The retention rate $r = \frac{F}{I}$  quantifies the efficiency of atom utilization from the initial loading, where $F$ is the final number of atoms and $I$ is the number of initially loaded atoms.

\textbf{Definition:} The \emph{rearrangement time} of the atoms includes both the computational time it takes to calculate the movement sequence and the physical time it takes to transport the atoms. Minimizing this time is essential to prevent decoherence and atom loss through background collisions.

As demonstrated in~\cite{tian_parallel_2023}, the physical time dominates the total duration. Therefore, the primary goal of rearrangement algorithms is to minimize the number of moves executed. In addition, the total distance should be minimized, which will lead to reducing the physical time~\cite{barredo_atom-by-atom_2016}. Despite the physical time dominating, the algorithm complexity and scalability become increasingly important as the initial lattice size increases.

\section{Methods}
\label{sec:methods}

ATLAS is a method to rearrange the loaded atoms to form a defect-free square-shaped sublattice $N$ of dimensions $L\times L$, where $L \leq W$. The rearrangement should follow the following physical constraints: 1) maximum acceleration $a_{max}=2750 m/s^2$, 2) trap transfer time $t_{transfer}=60 \mu s$, 3) maximum moving velocity $v_{max} = 0.13 m/s$, 4) spacing between SLM traps $d=5 \mu m$, 5) due to the use of AODs for shuttling, the relative order of rows and columns must be preserved during movement. ATLAS follows realistic experimental constraints and consists of two main phases.

\subsection{Planning Phase}

In the planning phase, a virtual copy of the loaded $W \times W$ lattice is used, assuming perfect transport (i.e., zero loss probability). All moves required to assemble a defect-free $L \times L$ target region are computed, but not yet executed. The output is an ordered list of parallel move batches which, when applied in sequence, guarantee a fully filled target block.

The planning phase consists of seven sequential subroutines. We use a \emph{movement simulator} to track the effect of the rearrangement. After each subroutine completes, the simulator’s movement history is copied into a list of planned moves and then cleared. This ensures that the execution phase (see next section) can replay the same movement batches in the correct order.

\begin{figure*}
    \centering
    \subfloat[]{\includegraphics[width=0.12\linewidth]{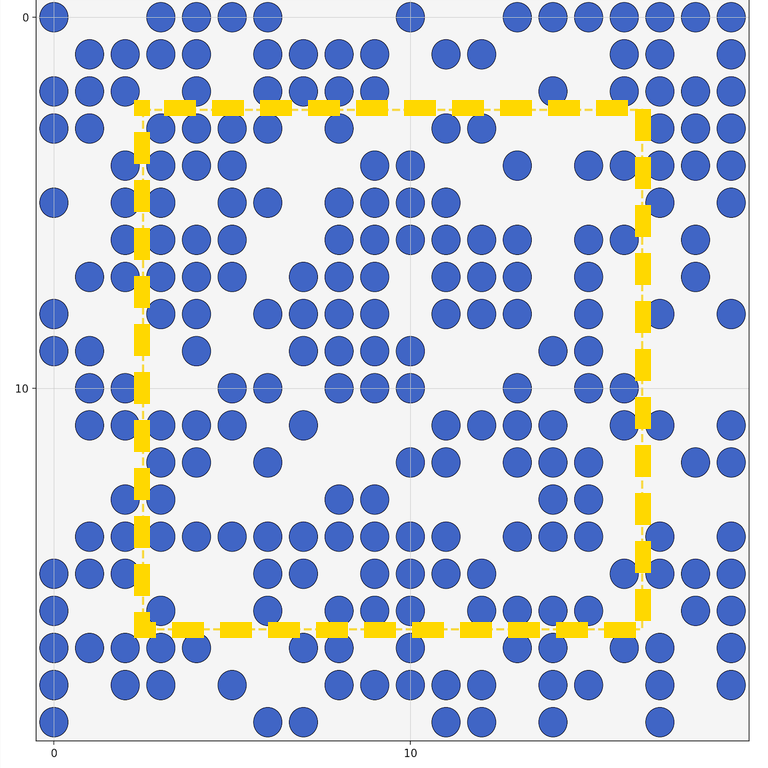}\label{fig:step_0}}
    \hfil
    \subfloat[]{\includegraphics[width=0.12\linewidth]{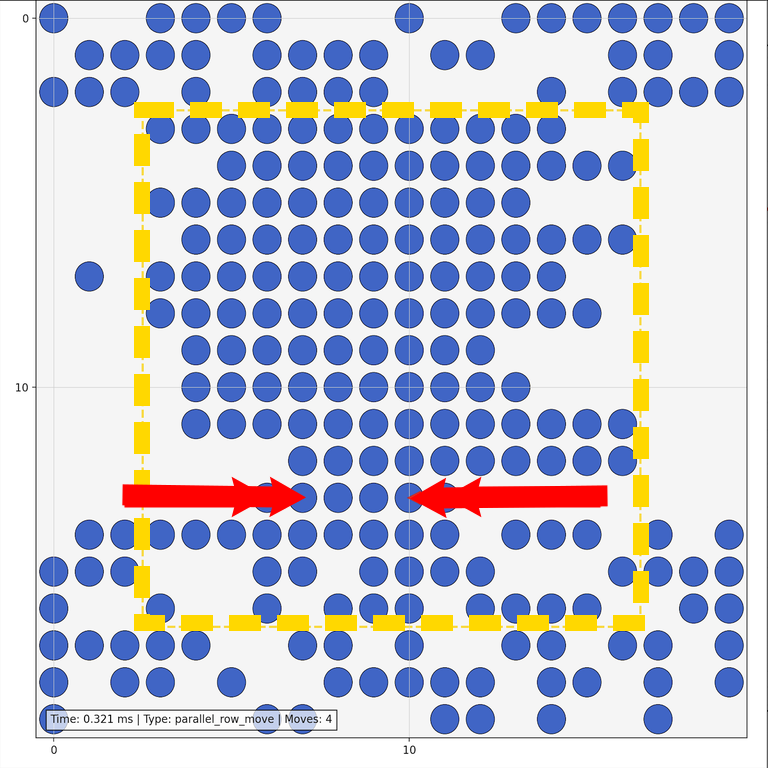}\label{fig:step_1}}
    \hfil
    \subfloat[]{\includegraphics[width=0.12\linewidth]{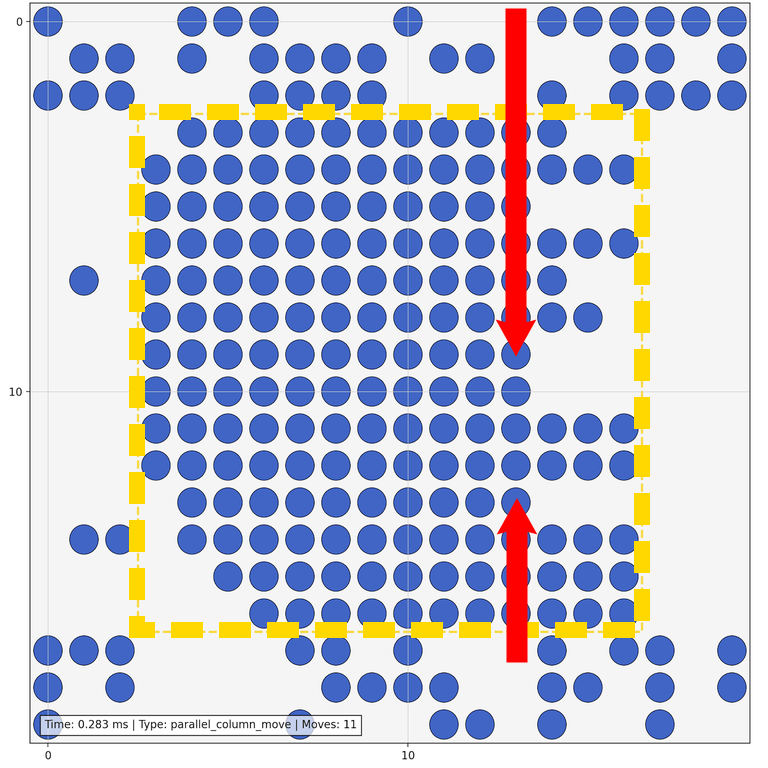}\label{fig:step_2}}
    \hfil
    \subfloat[]{\includegraphics[width=0.12\linewidth]{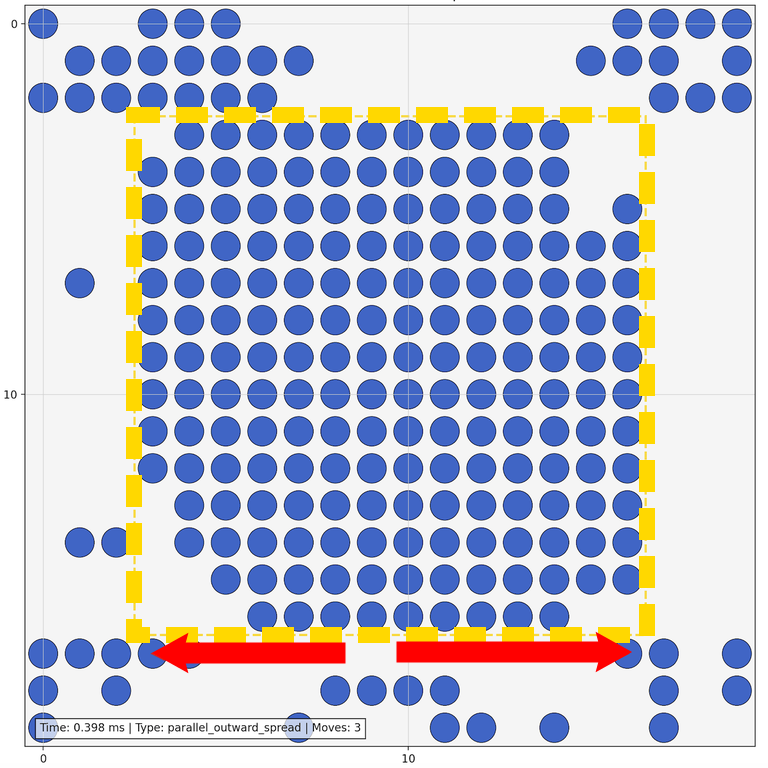}\label{fig:step_3}}
    \hfil
    \subfloat[]{\includegraphics[width=0.12\linewidth]{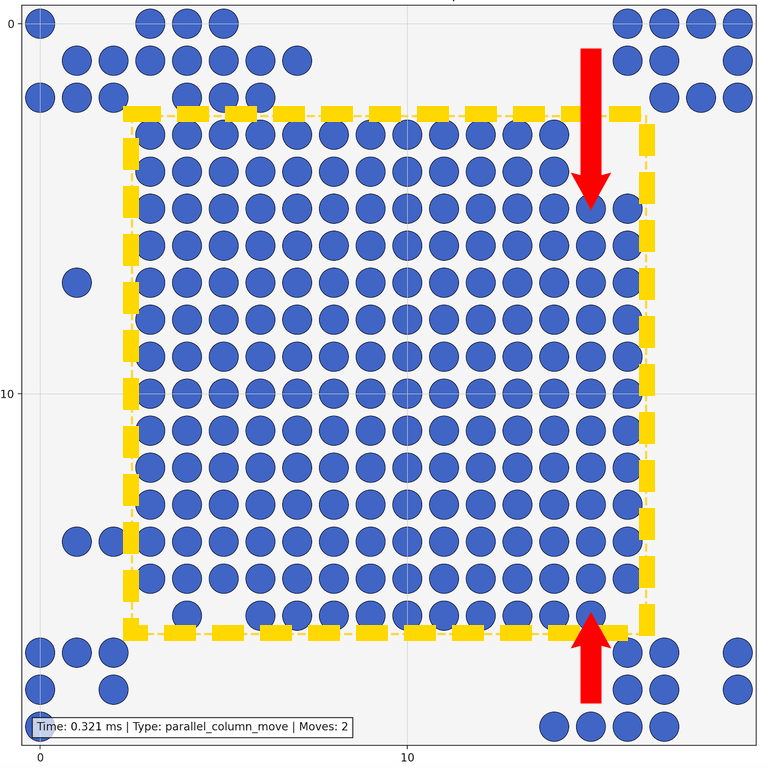}\label{fig:step_4}}
    \hfil
    \subfloat[]{\includegraphics[width=0.12\linewidth]{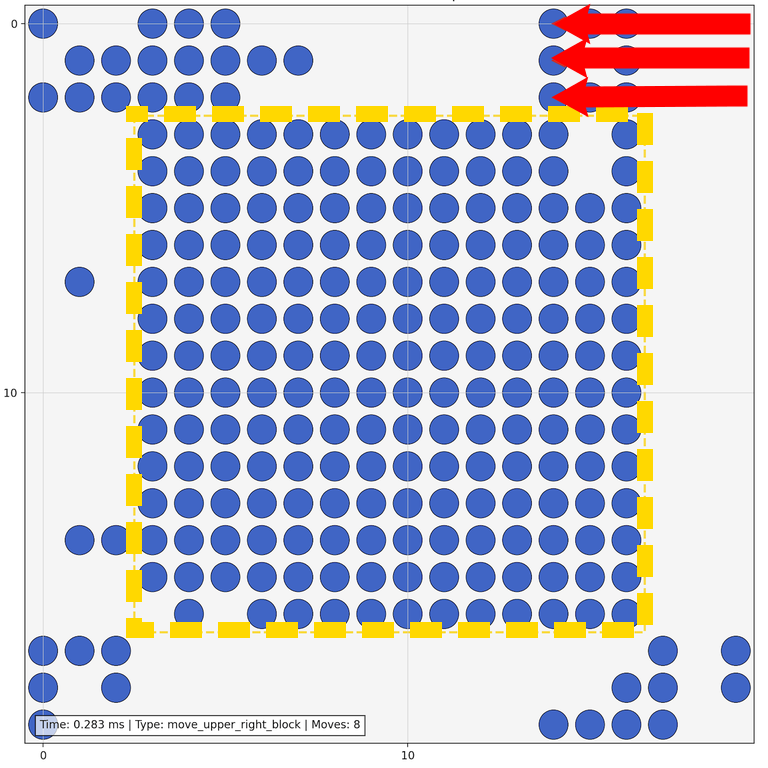}\label{fig:step_5}}
    \hfil
    \subfloat[]{\includegraphics[width=0.12\linewidth]{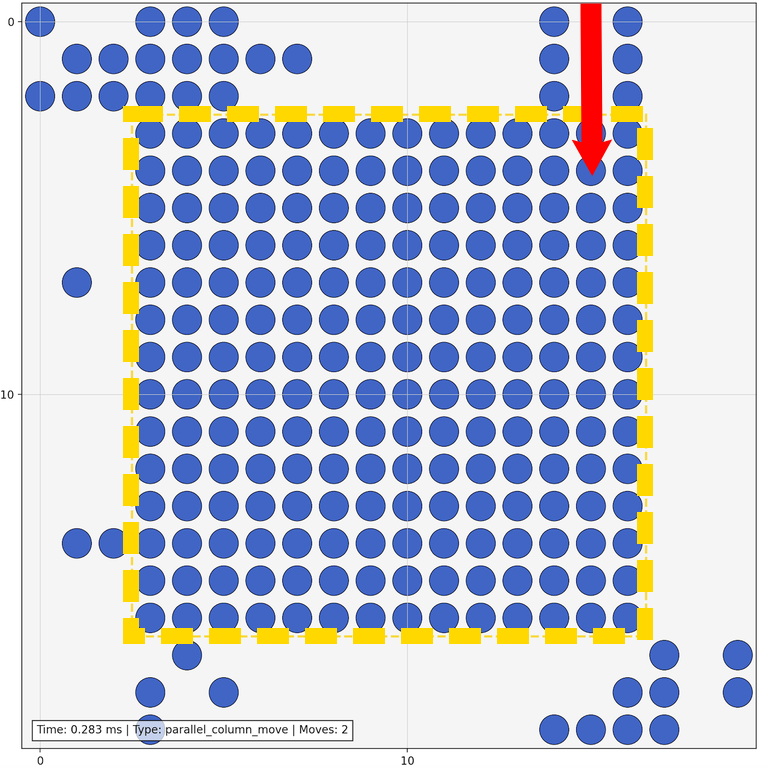}\label{fig:step_6}}
    \caption{ATLAS without loss ($p_{loss}=0$).
    (a) The atoms are initially loaded into each site with a 0.6 probability. The red arrows represent parallel moves made in one move batch.
    (b) Row-wise centering is performed, where each row fills its defects using atoms from the same row and side.
    (c) Column-wise centering starts from the left side of the target zone and progresses towards the right edge. Four move batches remain after the one shown.
    (d) The "spread" phase moves atoms above and below the target zone. Atoms on the right side are shifted to the right edge of the target zone, and atoms on the left side are moved toward the left edge.
    (e) The "squeeze" phase inserts atoms into the target zone after spreading.
    (f) Atoms in the top-right corner block are moved in parallel. The bottom-left corner is moved the same way immediately after.
    (g) A final column-wise centering "squeezes" the remaining atoms into the target zone. If any defects remain, additional steps are executed.}
    \label{fig:step_by_step}
\end{figure*}

\paragraph{1. Target‐Zone Initialization}  
We choose $L$ based on the initial atom budget $I=\sum_{i,j}M_{i,j}$, the per‐move loss probability $p_{loss}$, a reference size $W_{0}=30$, base move count $m_{0}=2$, and a safety margin $s=0.95$.  Concretely, we set
\[
    m = m_{0}\sqrt{\frac{W}{W_{0}}},
    \qquad
    I_{eff} = I(1-p_{loss})^{m}s,
    \qquad
    L = \lfloor\sqrt{I_{eff}}\rfloor. 
\]
We then center the $L\times L$ block in the $W\times W$ field by computing 
\[
    \delta=\lfloor\frac{W - L}{2}\rfloor
\]
and declare every site $(i,j)$ with $\delta \leq i < \delta + L$, $\delta \leq j < \delta + L$ to be in the target zone.  All subsequent routines test membership in this region in $O(1)$ time.

\paragraph{2. Row‐Wise Centering}

For each of the $L$ rows in the target region, we split the row at its center. On the left side, the defects are processed in order from the center out toward the left edge; each defect is paired with the closest atom to its left (which may lie outside the target block’s columns). On the right side, we process defects from the center toward the right edge, pairing each with the nearest atom to its right. All source$\rightarrow$target assignments for that row are then executed in parallel as a single batch.

\paragraph{3. Column‐Wise Centering}  
The same centering is applied to each of the $L$ columns of the target region, again batching all moves within each column.

\paragraph{4. Spread‐and‐Squeeze Cycle} consists of: \emph{Spread:}  In each row immediately above and below the target band, we "slide" any atom that lies horizontally within the band span outward (left on the left side, right on the right), collecting one move batch per row. \emph{Squeeze:}  We follow with another column‐wise centering to move those newly spread atoms into the target. We repeat this spread‐and‐squeeze cycle a fixed number of times (typically four) or until no further atoms can be moved.

\paragraph{5. Corner‐Block Moves}. Any atoms remaining in the four $\delta \times \delta$ corner regions are considered in unison.  We test whether all four corner blocks can shift inward by $\delta$ without collision; if so, we move them in one batch. Otherwise, we attempt upper/lower or left/right pairs, and finally any single corner block. After corner movement, we run one more column‐wise centering to potentially fix defects in the target zone.

\paragraph{6. Second Spread‐and‐Squeeze}  

A final, shorter spread‐and‐squeeze iteration (up to three cycles) ensures that any atoms brought into the edges by corner moves are fully drawn into the central zone.

\paragraph{7. Defect Repair}  

Any remaining defects are then repaired individually. For each defect, we select the nearest available atom outside the target band—using direct, L‐shaped, or A* pathfinding—and transport it along a collision‐free path. These moves are necessarily sequential, i.e., moving one atom at a time. In practice, however, this phase is rarely needed since the defects are usually filled by the earlier subroutines.

At the end of the planning phase, the virtual lattice is guaranteed to be perfect and we have a complete, ordered list of parallel‐move batches ready for execution.

\subsection{Execution Phase}

We \emph{reset the movement simulator} to an imperfect field, restore the true atom‐loss probability $p_{loss}$, and replay each batch computed before. For each batch:
\begin{enumerate}
  \item We filter out any moves whose source atom has been lost.
  \item We compute the longest Manhattan distance in the batch and use our trapezoidal velocity model (plus fixed trap‐transfer delays) to assign a physical movement time.
  \item We apply loss probabilistically to each move, updating the lattice and recording which transports succeeded or failed.
\end{enumerate}

Each atom movement has a probability $p_{loss}$ of losing the atom completely. For example, if one atom has a path that is L-shape with $p_{loss}=0.05$, the whole probability of surviving the path is $(1-p_{loss})^2=(1-0.05)^2=0.95^2=0.9025$.

Because most heavy computation (pathfinding, blockage checks, greedy matching) occurs in planning, the execution phase is lightweight. On lattices up to $100 \times 100$, the full two-phase strategy completes quickly on standard hardware. Physical movement time is primarily determined by the number of move batches, rather than the individual distances involved.

\section{Results}
\label{sec:res}

ATLAS was implemented in Python 3.11.4, using NumPy 1.26.4 for numerical operations and Matplotlib 3.10.1 for visualization. Source code is available at [removed for review]. All experiments and performance measurements presented were carried out on a 2021 MacBook Air (M1) with 16 GB of RAM and a 256 GB SSD.

The algorithm is tested with different initial lattice sizes ($10 \times 10$, $20 \times20$, $50\times50$, $75\times75$, and $100\times 100$), load probabilities (0.5, 0.7, and 0.9), and atom loss probabilities (0.0, 0.01, and 0.05) using Monte Carlo simulation. Each configuration is simulated 100 times with random seeds from 0 to 99. Each simulation terminates either when a perfect fill rate is achieved or after six iterations. The six-iteration limit is chosen to reduce simulation time.

\subsection{Physical Model, Movement Timing, and Loss}

To approximate realistic atom transport, each move is modeled with a trapezoidal velocity profile. A maximum acceleration of $a_{\max}=2750$ m/s$^2$ is assumed until the atom reaches $v_{max}=0.13$ m/s, followed by a symmetric deceleration. The time to "pick up" or "drop off" an atom at a static trap is fixed at $60 \mu$s each, so the total physical time for a move of Manhattan distance $d$ (in lattice‐site units) is$
    t_{move}(d)
    = 2 \times 60\mu s
    + t_{kin}(d)$, where
where $t_{kin}(d)$ is the kinematic time computed as follows.

Let $D = d \times d_{site}$ denote the physical distance (with $d_{site}=5 \mu$m). Define  $D_{accel} = \frac{v_{\max}^2}{2a_{\max}}$ as the distance required to accelerate to top speed. If $2D_{accel}\leq D$, then $
    t_{kin}(d)
    = 2\frac{v_{max}}{a_{max}}
    +\frac{D - 2D_{accel}}{v_{max}}$, corresponding to accelerate–coast–decelerate. Otherwise
$t_{kin}(d)= 2\sqrt{\frac{D}{a_{max}}}$, a purely triangular accelerate–decelerate profile.

During the execution phase, each individual atom transfer carries a probability $p_{loss}$ of losing the atom entirely. This is modeled by sampling a Bernoulli trial for each move in a batch: successful moves update the atom’s position in the lattice, while failed moves zero out the source cell and leave the destination empty. All successes and failures are recorded in the move batch’s summary.

\subsection{Fill Rate Analysis}
\label{sec:fill}

Since each Monte Carlo trial terminates upon first reaching $\eta=1$ or after six relocation iterations, the mean fill rate $\langle \eta\rangle$ captures both the algorithm’s converged success and its residual defects under finite iterations and stochastic loss.

\begin{figure*}
    \centering
    \subfloat[]{
        \includegraphics[width=0.49\linewidth]{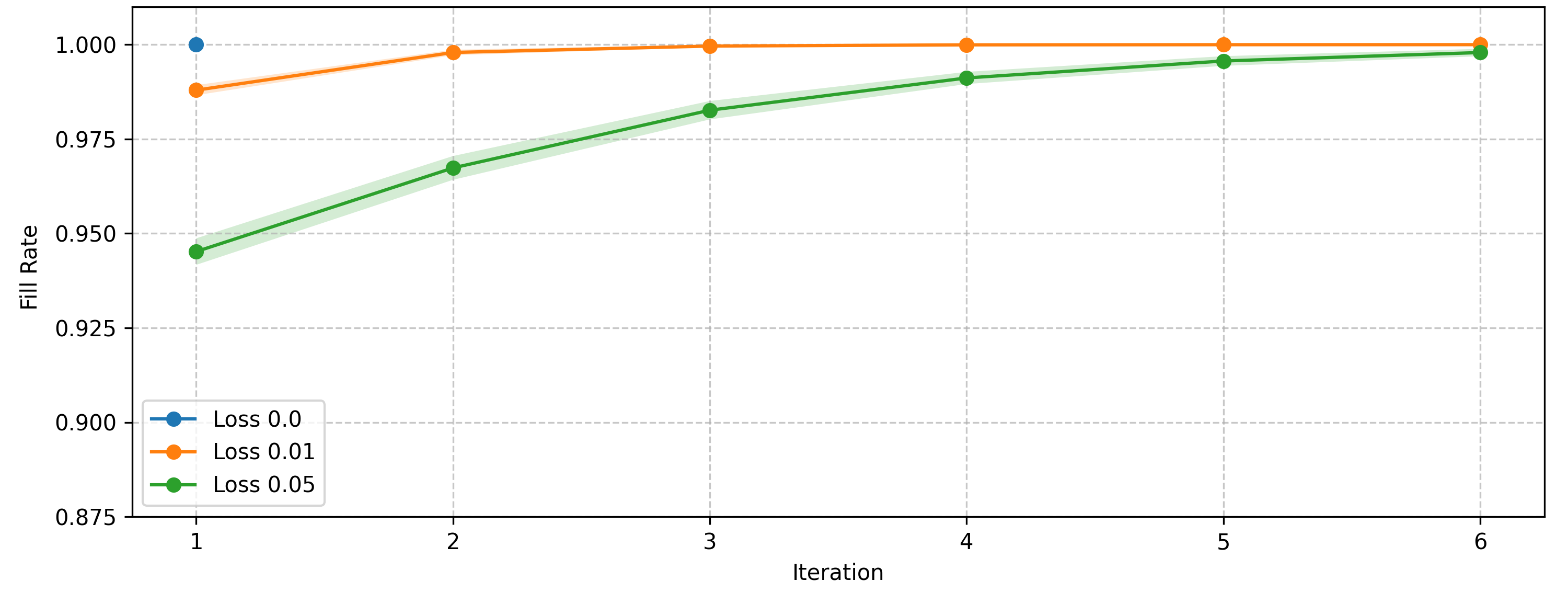}
        \label{fig: fillrate_vs_iteration}}
    \subfloat[]{
        \includegraphics[width=0.49\linewidth]{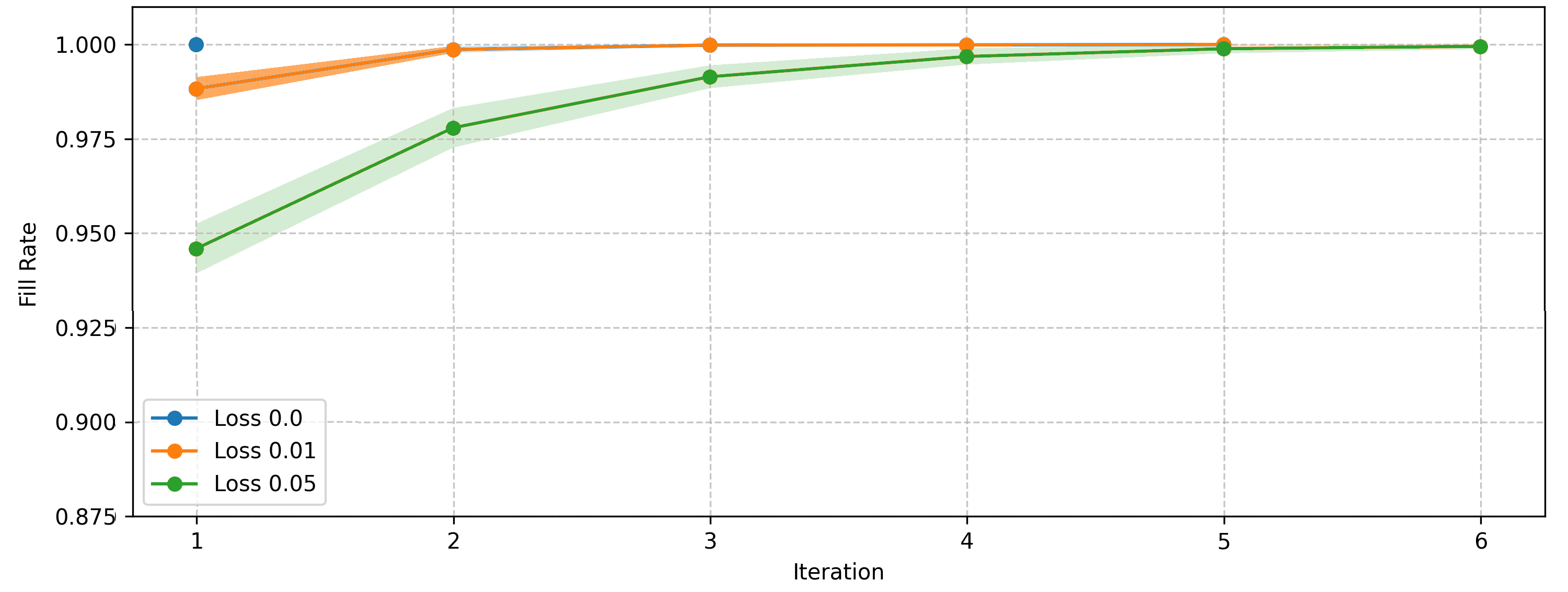}   
        \label{fig: fillrate_vs_iteration50}
    }
    \caption{Monte Carlo simulation of fill rate improvement over iterations for: (a) a $100 \times 100$ lattice with an occupation probability of 0.7; (b)  a $50 \times 50$ lattice with an occupation probability of 0.7. With $p_{loss} = 0$ (blue), the algorithm achieves a perfect fill rate immediately in the first iteration.}
\end{figure*}

The data shows that ATLAS consistently achieves high fill rates (typically > 99\% within 6 iterations) with low standard deviation, indicating strong reliability. Fig.~\ref{fig: fillrate_vs_iteration} and~\ref{fig: fillrate_vs_iteration50} show the fill rate evolution across iterations for $100 \times 100$ and $50 \times 50$ lattices, respectively, with an occupation probability of 0.7. In both cases, the fill rate converges toward 1, although more iterations are required as the atom loss probability increases. The similarity in convergence behavior across lattice sizes highlights the scalability of the approach.

\begin{figure}
    \centering
    \includegraphics[width=0.7\linewidth]{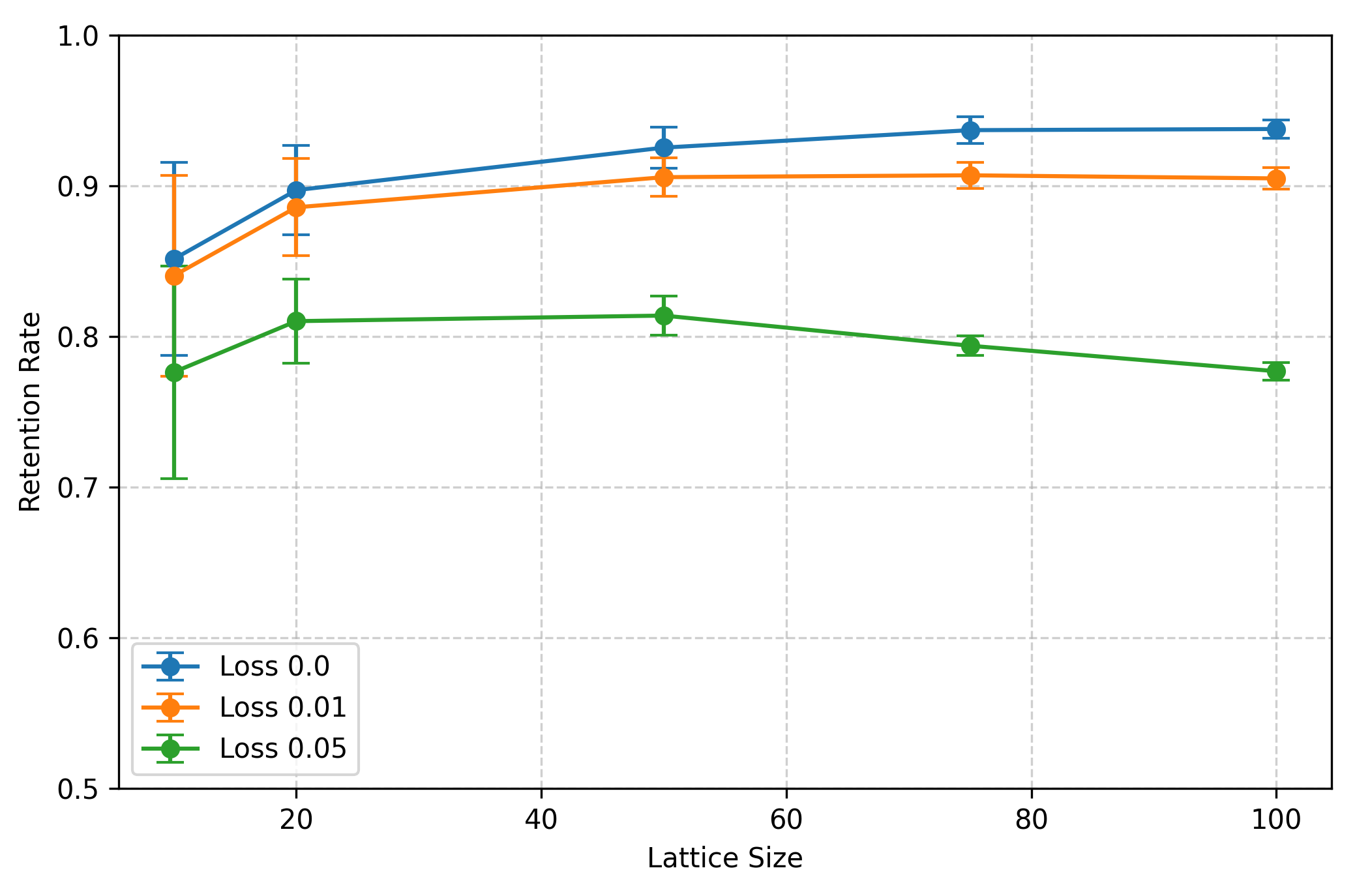}
    \caption{Monte Carlo simulations of retention rate as a function of initial lattice size, shown for three different atom-loss probabilities; blue: $p_{loss}=0$, orange: $p_{loss}=0.01$, and green: $p_{loss}=0.05$. The retention rate decreases with increasing loss probability. The more significant drop for $p_{loss}=0.05$ as the lattice size increases is an artifact of capped iterations. Without the cap, retention would remain near 80 \%.}
    \label{fig:retentionrate_vs_size}
\end{figure}

For a $50 \times 50$ lattice with an occupation probability of 0.7, a perfect fill rate is consistently achieved after the first iteration in the absence of loss. This trend is consistent across all lattice sizes and occupation probabilities.

 With a moderate loss ($p_{loss}=0.01$) convergence requires on average $\approx 2.970$ iterations, with a standard deviation of 0.674. With even higher loss ($p_{loss}=0.05$), the average number of iterations increases to $\approx 5.326$ with a standard deviation of 0.659. Although the number of iterations increases, the mean fill rate remains perfect. However, with $p_{loss}=0.05$ and occupation probabilities of 0.5 and 0.7, a small standard deviation in fill rate (0.001) appears, indicating occasional runs that reach the iteration cap with a few sites remaining unoccupied.

Even with a $100 \times 100$ lattice, the mean fill rate remains high. Only under $p_{loss}=0.05$ does the mean fill rate slightly decrease from perfect to 0.998 with a standard deviation of 0.001. That occurs because most runs reach the iteration cap before all defects are resolved. With a higher iteration limit or without the cap at all, the mean fill rate is expected to reach 1.0 as well.

\subsection{Retention Rate Analysis}
\label{sec:ret}

Fig.~\ref{fig:retentionrate_vs_size} shows Monte Carlo results at a fixed occupation probability $p_{occ}=0.7$ and three different loss probabilities. Fig.~\ref{fig:retentionrate_vs_size} demonstrates that the retention rate is generally lowest for the smallest lattice sizes, suggesting that the algorithm is less efficient for these configurations. This is due to how the target zone is initialized. In small lattices, the safety margin ($s=0.95$) applied during the sizing of the target zone has a bigger impact than in larger lattices. For instance, shrinking the target zone from  $7 \times 7$ to $6 \times 6$ will reduce the number of atoms by over 26 \%. For larger lattices, the impact of the safety margin is not as significant, and the retention rate increases and eventually plateaus.

The trend is similar to Fig.~\ref{fig:retentionrate_vs_size} across different occupation probabilities. However, when excluding the smallest lattice size ($10 \times 10$), the retention rate improves slightly with increasing occupation probability. The highest average retention rate happens at $p_{occ}=0.9$ and the lowest at $p_{occ}=0.5$. The difference between the two is approximately 0.01, which is relatively small.

\subsection{Efficiency Analysis}
\label{sec:time}

\begin{figure}
    \centering
    \includegraphics[width=0.6\linewidth]{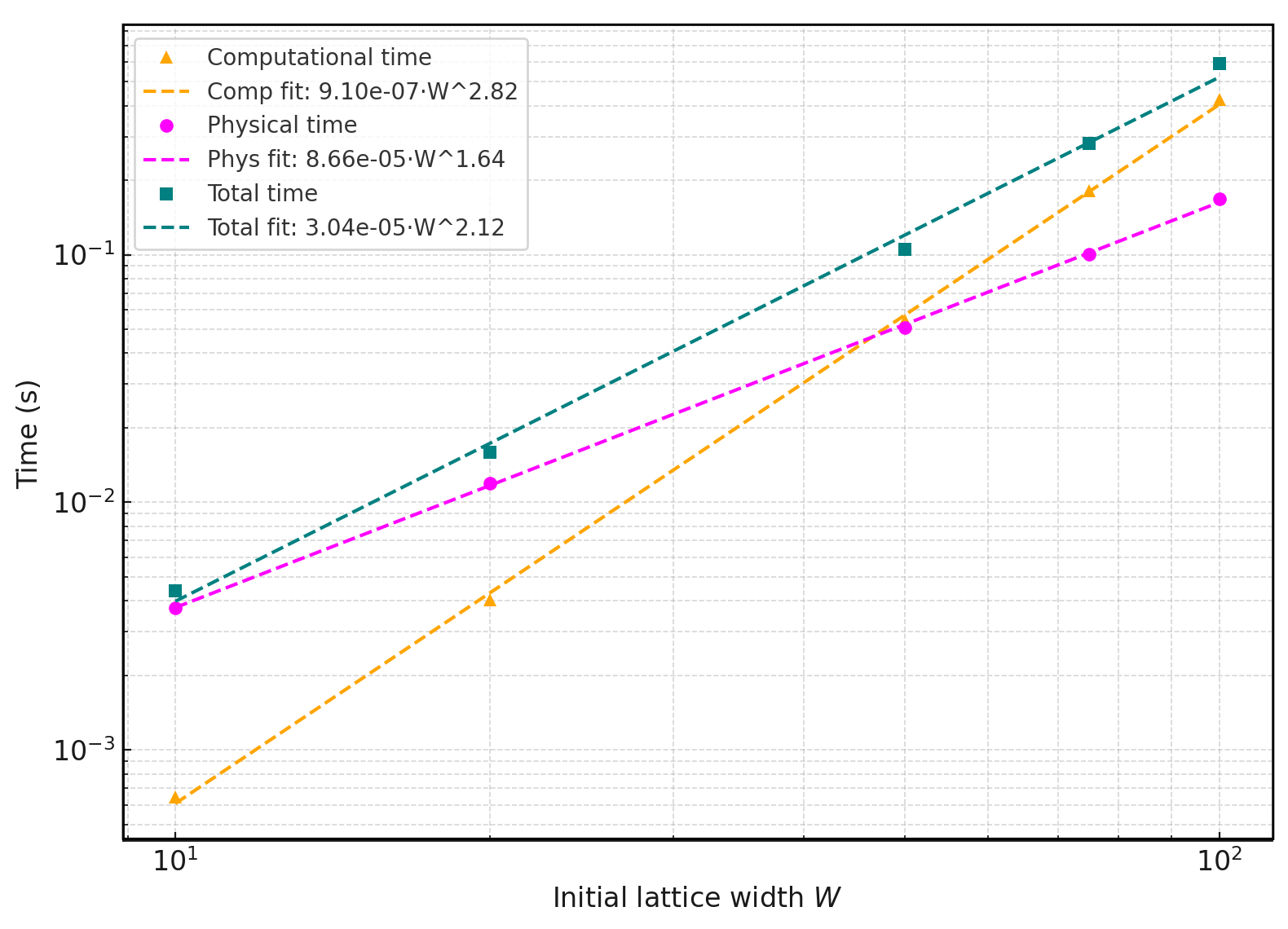}
    \caption{Log–Log Scaling of Rearrangement Time vs. Lattice Width:
    Monte Carlo–measured timings plotted on a log-log scale as a function of lattice side length W (for $p_{loss}=0$). The yellow line represents the computational time scaling with the initial lattice width. The magenta line represents the physical time it takes to complete the moves. The green line represents the total time (computational + physical). The yellow line shows nearly cubic scaling of computational time with lattice width. Above $W\approx50$, the computational time starts to dominate the total time.}
    \label{fig:time_vs_size}
\end{figure}

The computational time scales approximately cubically with the initial lattice width. In our current data, this computational time is the dominating factor in the total execution time, Fig.~\ref{fig:time_vs_size}. However, the hardware we use is significantly less efficient for this task than the supercomputers, where this algorithm would run in real-world simulations. On such high-performance computing platforms, the total time is dominated by physical time, i.e., the time it takes to move the atoms. Overall, the results suggest that the move count scales sublinearly with the number of sites, which indicates good scalability.

\begin{figure*}
    \subfloat[]{
    \includegraphics[width=0.33\linewidth]{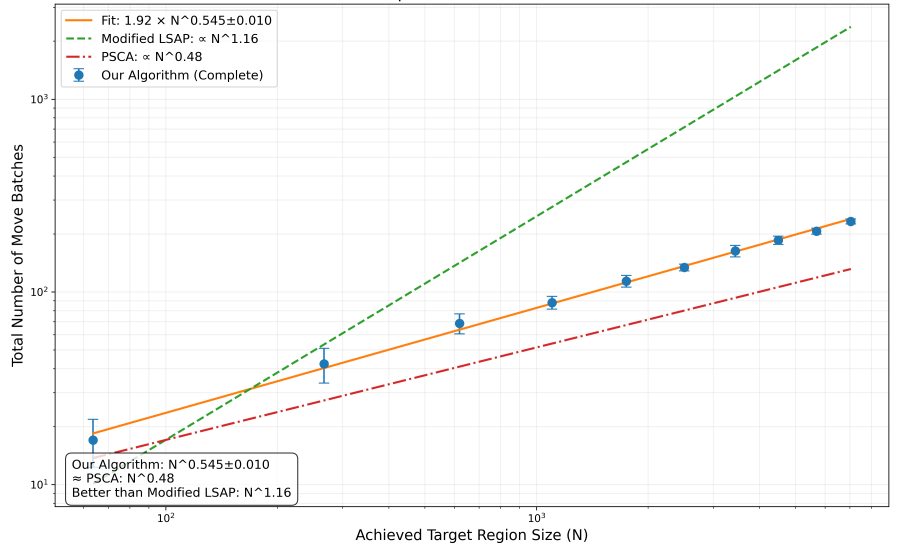}
    \label{fig:0.75_comparison}
    }
    \subfloat[]{
    \includegraphics[width=0.37\linewidth]{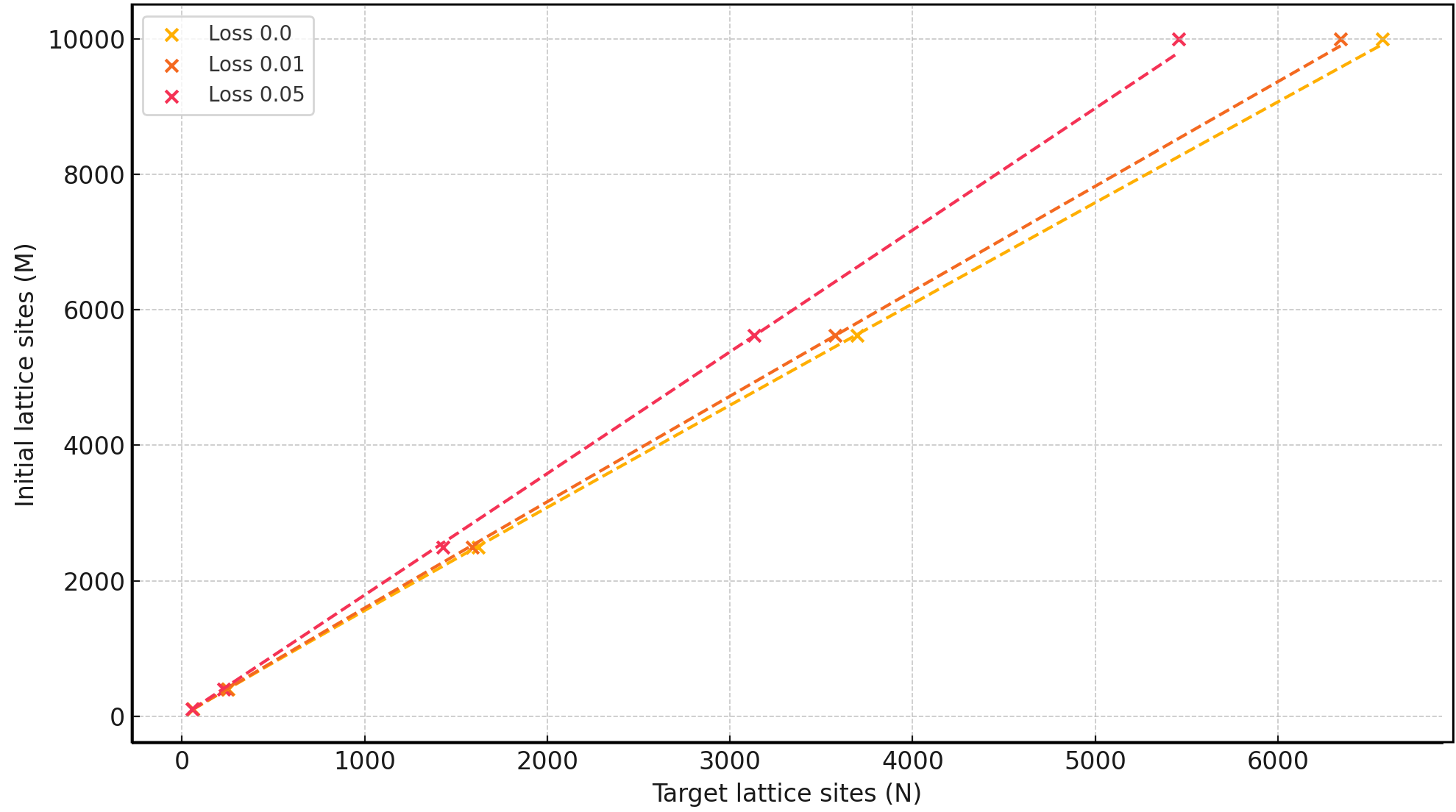}
    \label{fig: M_function_N}
    }
    \subfloat[]{
    \includegraphics[width=0.295\linewidth]{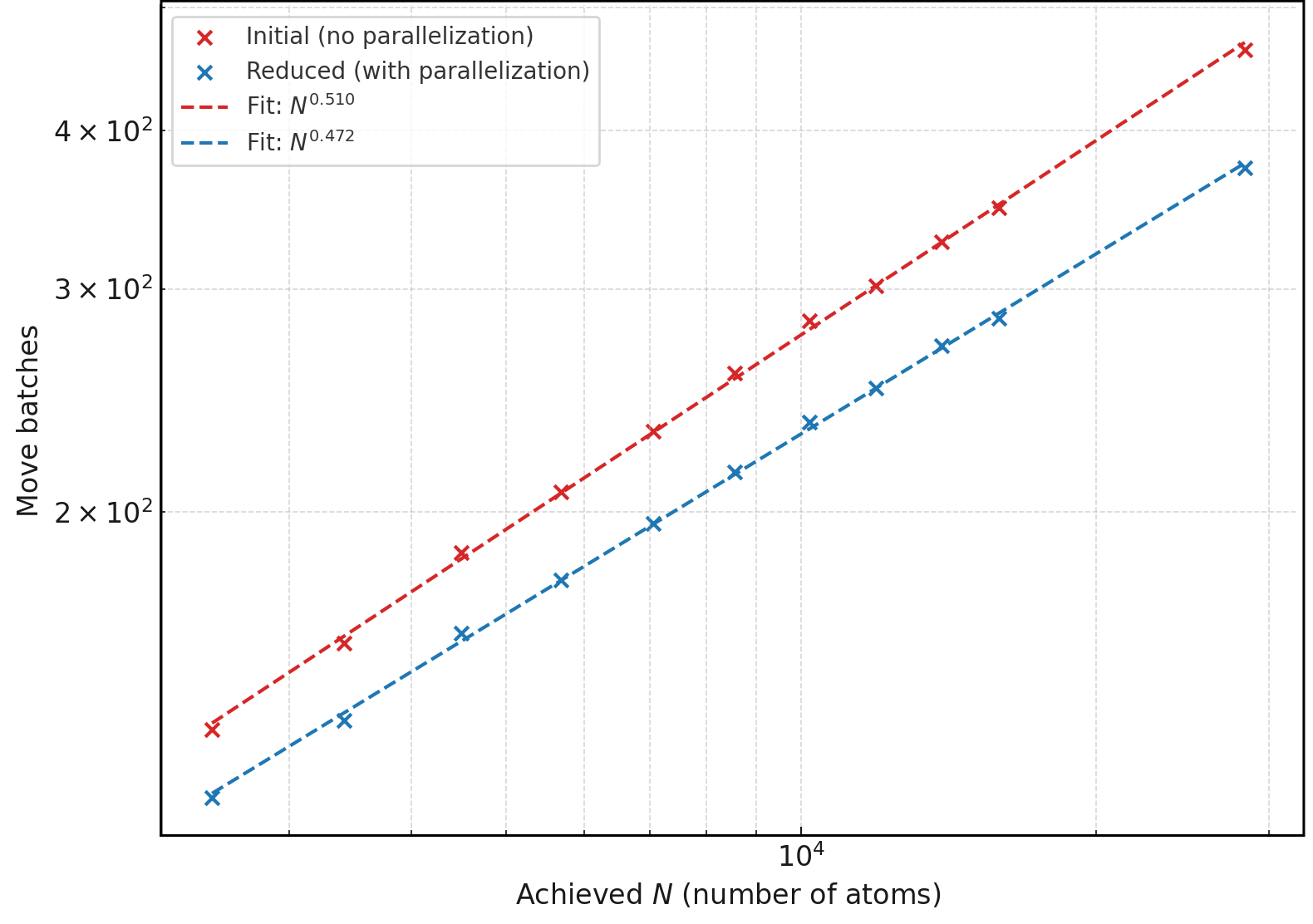}
    \label{fig:afterParallelization}
    }
    \caption{
    (a) Moves scaling as a function of target lattice size. Modified LSAP (single-tweezer) algorithm shows slightly above linear scaling $N^{1.16}$ (green line). Non-parallel ATLAS scales $\approx N^{0.545}$ (orange line and blue dots). Whereas, PSCA shows the scaling of $N^{0.48}$ (red line).
    (b) Initial lattice size scaling with target size. The occupation probability is 0.7, red line represents $p_{loss}=0.05$, orange $p_{loss}=0.01$, and  yellow $p_{loss}=0$. All three curves exhibit linear scaling ($M\propto N$), even at a 0.05 loss probability.
    (c) Moves scaling as a function of target lattice size ($L=60-200$). Our previously used algorithm (red line) shows worse scaling $N^{0.510}$ than PSCA, but using this parallelization method, we achieve $N^{0.472}$ vs. PSCA's $N^{0.48}$.}
\end{figure*}

\begin{figure}
    \centering    
    \includegraphics[width=0.6\linewidth]{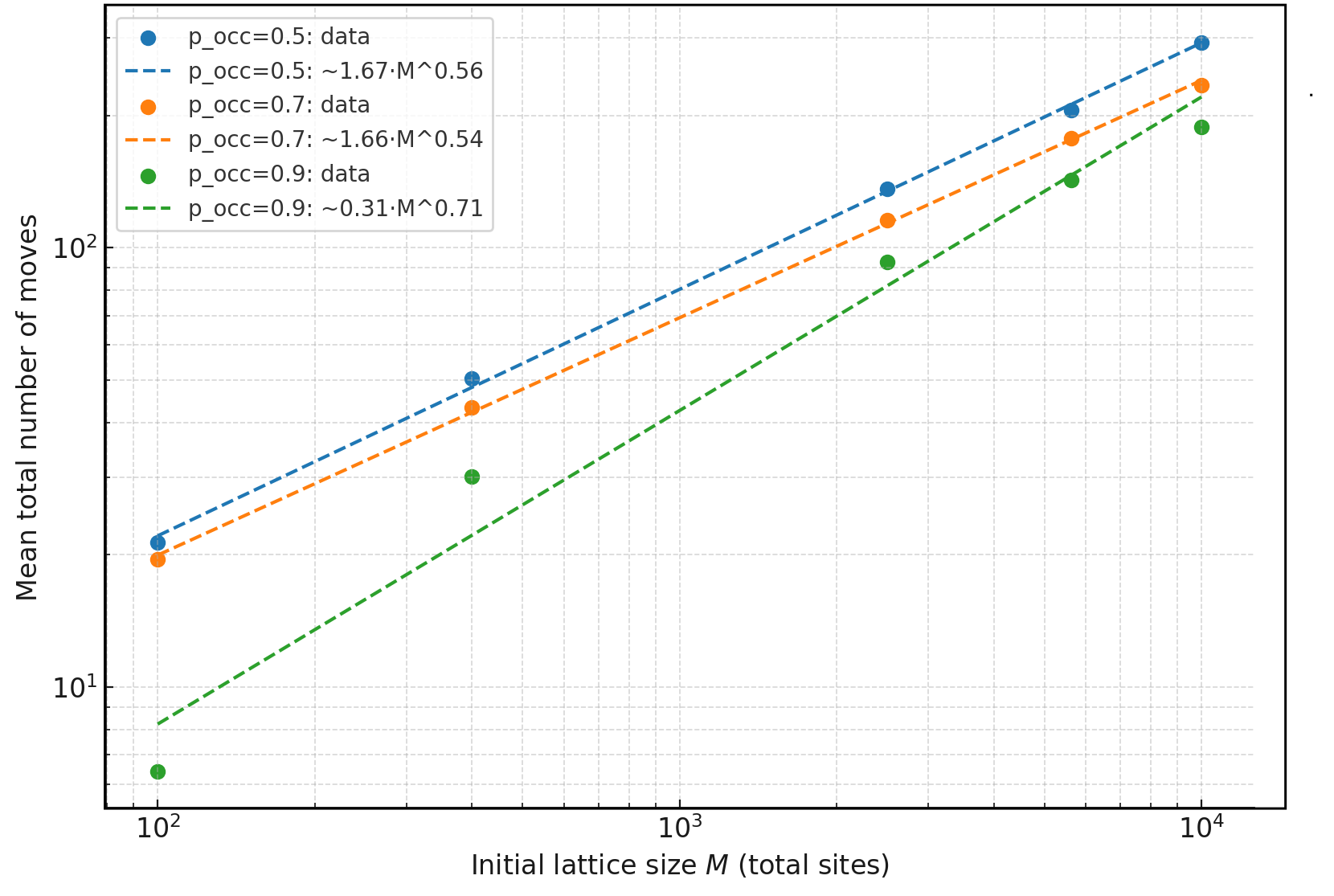}
    \caption{Monte Carlo–computed scaling of rearrangement moves as a function of total sites $M$, at zero atom-loss probability. Moves scaling for three occupation fractions: $p_{occ}=0.5$, 0.7, and 0.9.  For $p_{occ}=0.5$ and 0.7, the exponent remains $\approx0.55$, while at high fill ($0.9$) the full-range fit yields $\approx0.71$.
    }
    \label{fig: efficiency_scaling}
\end{figure}

In Fig.~\ref{fig: Moves_w_diff_occProb} the move count scales approximately as $M^{0.55}$ with occupation probabilities of 0.5 and 0.7, where $M=W \times W$ is the total number of trap sites. At a higher occupancy of 0.9, the exponent increases to 0.71. However, the plot fitting is highly affected by the smaller lattice sizes. When considering only larger lattices ($M \geq 2500)$, the exponent drops to 0.51, aligning with the lower-occupancy cases. This suggests that the scaling is not affected by the occupation probability.

\subsection{Comparison with existing algorithms}

Existing algorithms are commonly benchmarked by analyzing how the number of moves scales with target lattice size $N$~\cite{tian_parallel_2023,wang_accelerating_2023,cimring_efficient_2023}.

Fig.~\ref{fig:0.75_comparison} shows that ATLAS achieves $N^{0.545}$, substantially better than the single-tweezer Modified LSAP algorithm~\cite{knottnerus_parallel_2025}, which scales as $N^{1.16}$. The multi-tweezer algorithm PSCA~\cite{tian_parallel_2023} slightly outperforms us at $N^{0.48}$ when $p_{occ}=0.75$.

However, the moves scaling alone does not capture the full picture: retention rate critically affects the efficiency. The PSCA begins by loading $20 \times 20$ static tweezers at 75–78 \% efficiency. Following that, they demonstrated the rearranging being dominated by physical movement by constructing a defect-free array of size $15 \times 15$ from an initial array of $20 \times 20$.

Based on this demonstration, and assuming a loading efficiency of $\approx 0.75 \%$, we estimate the retention rate of PSCA as:
\[
    r_{PSCA} = \frac{F}{I}
    \approx \frac{15\times15}{(20 \times 20) \times 0.75} 
    = \frac{225}{300}
    =0.75    
\]

By contrast, on the same  $20 \times 20$ initial grid ATLAS achieves a mean target of 267 atoms ($r\approx 0.89$). It corresponds to usually forming a $16 \times 16$ or $17 \times 17$ target lattices. Even increasing the target size from $15 \times 15$ to $16 \times 16$ yields 31 additional qubits. Moreover, even when introducing an atom loss of 0.05, ATLAS maintains a retention rate of $r \approx 0.8$, still exceeding PSCA's performance.

Moreover, the required initial lattice size $M$ to achieve a target $N$ remains essentially linear in ATLAS, even under loss (Fig.~\ref{fig: M_function_N}). This outperforms the approach in~\cite{cimring_efficient_2023}, which also scales linearly in the absence of loss, but degrades to $M \propto N^{3/2}$ when a loss is introduced.
This linear scaling not only reduces hardware overhead but also lowers demands on laser power and control channels, making the preparation of even larger arrays experimentally feasible.

\subsection{Parallel ATLAS}
\label{sec:par}

We reduced the scaling exponent of the moves from 0.545 (this exponent is highly affected by the small lattice sizes) to 0.472 in the absence of loss. First, we we iterate each step until no further improvement is made.

Second, the algorithm still simulates every move in the simulated lattice and collects the movement history. Then one by one, we check if moves can be combined into a single batch by following the six rules made to follow AOD restrictions: 1) \textbf{Static atom blocking:} A move cannot pass through a static atom on the same row or column. 2) \textbf{No crossing active lines:} A move may not cross a static atom on another simultaneously active line, though it may end aligned with it. 3) \textbf{Unique endpoints:} Parallel moves cannot share the same start or end position. 4) \textbf{No path intersection:} Moves whose paths overlap or intersect cannot run in parallel. 5) \textbf{Ordering within a line:} Moves sharing a row (or column) must preserve left–right (or top–bottom) order after the move. 6) \textbf{Ordering across lines:} Moves on different rows must maintain relative column order, and moves on different columns must maintain relative row order; shared rows/columns must also share consistent endpoints.

We will compare each collected move and check if they can be combined. The algorithm compares the first move first with everyone, then the second, and so on. This reduces the moves significantly, especially in the larger lattices. With parallelization, the number of move batches is scaling even better than they had in PSCA. Because ATLAS is trying to have as many atoms as possible in the final lattice, it will not perform well with smaller lattices. This is why we show the moves scaling using larger lattices $L=60$ to $L=200$ in Fig.~\ref{fig:afterParallelization}.

\section{Conclusion}

We presented \textsc{ATLAS}, an efficient and loss-aware atom rearrangement
algorithm for assembling defect-free neutral-atom arrays under realistic
experimental constraints. By separating planning and execution, exploiting
extensive parallelism, and incorporating transport loss into the target-size
estimation, ATLAS reliably produces high fill rates and strong retention across
a wide range of lattice sizes, loading probabilities, and loss conditions.
Monte Carlo simulations demonstrate sublinear move scaling, linear initial-size
requirements, and high robustness even at nonzero loss, outperforming existing
approaches in both scalability and atom utilization. A parallelization refinement further reduces the move-scaling exponent to approximately $N^{0.47}$, comparable to the best reported multitweezer methods while achieving higher retention.

\balance
\bibliographystyle{ACM-Reference-Format}
\bibliography{__main}

\end{document}